\documentstyle[12pt]{article}

\begin{document}

\def\dg{^{\dagger}}
\def\al{\alpha}
\def\bt{\beta}
\def\gm{\gamma}        \def\Gm{\Gamma}
\def\dl{\delta}        \def\Dl{\Delta}
\def\ep{\epsilon}
\def\kp{\kappa}
\def\lm{\lambda}        \def\Lm{\Lambda}

\def\th{\theta}        \def\Th{\Theta}
\def\vph{\varphi}

\def\om{\omega}        \def\Om{\Omega}
\def\sg{\sigma}        \def\Sg{\Sigma}

\def\zt{\zeta}
\def\ie{{\it i.e.}}
\def\eg{{\it e.g.}}
\def\cf{{\it c.f.}}
\def\etal{{\it et.al.}}
\def\etc{{\it etc.}}
\def\np#1#2#3{Nucl. Phys. {\bf B#1} (#2) #3}
\def\pl#1#2#3{Phys. Lett. {\bf B#1} (#2) #3}
\def\plb#1#2#3{Phys. Lett. {\bf #1B} (#2) #3}
\def\prl#1#2#3{Phys. Rev. Lett. {\bf #1} (#2) #3}
\def\physrev#1#2#3{Phys. Rev. {\bf D#1} (#2) #3}
\def\prd#1#2#3{Phys. Rev. {\bf D#1} (#2) #3}

\def\nsp{{$NS5^\prime$}}
\def\nspp{{NS5^\prime}}
\def\gs{g_s}
\def\ddp{$Dp$}

\def\AA{{\cal A}}
\def\BB{{\cal B}}
\def\CC{{\cal C}}
\def\DD{{\cal D}}
\def\EE{{\cal E}}
\def\FF{{\cal F}}
\def\GG{{\cal G}}
\def\HH{{\cal H}}
\def\II{{\cal I}}
\def\JJ{{\cal J}}
\def\KK{{\cal K}}
\def\LL{{\cal L}}
\def\MM{{\cal M}}
\def\NN{{\cal N}}
\def\OO{{\cal O}}
\def\PP{{\cal P}}
\def\QQ{{\cal Q}}
\def\RR{{\cal R}}
\def\SS{{\cal S}}
\def\TT{{\cal T}}
\def\UU{{\cal U}}
\def\VV{{\cal V}}
\def\WW{{\cal W}}
\def\XX{{\cal X}}
\def\YY{{\cal Y}}
\def\ZZ{{\cal Z}}

\def\nf{N_f}
\def\nc{N_c}

\newcommand{\tr}{{\rm tr \,}}
\newcommand{\beq}{\begin{equation}}
\newcommand{\fre}{F[\vec{e},\vec{m}]}
\newcommand{\eeq}{\end{equation}}
\newcommand{\bea}{\begin{eqnarray}}
\newcommand{\eea}{\end{eqnarray}}
\newcommand{\dirac}{/\!\!\!\partial}
\newcommand{\Dirac}{/\!\!\!\!D}

\begin{titlepage}
\begin{center}
\hfill hep-th/9801020
\vskip .01in \hfill RI-11-97, EFI-97-59
\vskip .4in
{\Large\bf Branes, Orientifolds and Chiral Gauge Theories}
\end{center}
\vskip .4in
\begin{center}
{\large Shmuel Elitzur}${}^1$, {\large Amit Giveon}${}^1$,
{\large David Kutasov}${}^2$,
{\large David Tsabar}${}^1$
\vskip .1in
${}^1$ {\em Racah Institute of
Physics, The Hebrew University\\ Jerusalem 91904, Israel}
\vskip .05in
${}^2$ {\em EFI and Department of Physics, University of
Chicago\\ 5640 S. Ellis Av., Chicago, IL 60637 USA}
\vskip .02in
and
\vskip .02in
{\em Department of Particle Physics, Weizmann Institute of Science\\
Rehovot 71600, Israel}
\vskip .05in

\end{center}
\vskip .4in
\begin{center} {\bf ABSTR\,ACT} \end{center}
\begin{quotation}
\noindent
We discuss some aspects of the physics of branes in
the presence of orientifolds and the corresponding
worldvolume gauge dynamics. We show that at strong
coupling orientifolds sometimes turn into bound
states of orientifolds and branes, and give a
worldsheet argument for the flip of the sign
of an orientifold plane split into two disconnected
parts by an NS fivebrane. We also describe the moduli
space of vacua of $N=2$ supersymmetric gauge theories
with symplectic and orthogonal gauge groups, and analyze
a set of four dimensional $N=1$ supersymmetric gauge
theories with chiral matter content using branes.
\end{quotation}
\vfill
\end{titlepage}

\pagebreak
\section {Introduction and Discussion}
\label {sec-int}
In the last year important new insights were obtained
into the classical and quantum vacuum structure of
supersymmetric gauge theories in different dimensions
with various numbers of supersymmetries. This was
achieved by realizing the gauge theories in
question as low energy worldvolume theories
on branes in string theory (see \cite{gk} for
a review and references), and studying the
resulting brane configurations.
Some of the results that were found are:

\begin{enumerate}

\item Nahm's construction of the moduli
space of magnetic monopoles has been obtained
by using the description of monopoles
as $D$-strings
stretched between parallel $D3$-branes
in type IIB string theory \cite{diac}.

\item Montonen and Olive's
electric-magnetic duality in four
dimensional $N=4$ SUSY gauge theory as well
as Intriligator and Seiberg's mirror symmetry
in three dimensional $N=4$ SUSY gauge theory
were shown to be consequences of the non-perturbative
S-duality symmetry of type IIB string theory
\cite{tseytlin,gr,hanany}.

\item The auxiliary Riemann surface
whose complex structure was proven by
Seiberg and Witten to determine the low
energy coupling matrix of four dimensional
$N=2$ SUSY gauge theory was shown in
\cite{klmvw,witten} to be part of the
worldvolume of a fivebrane. Hence it
is physical in string theory.

\item Seiberg's infrared equivalence between
different four dimensional $N=1$ supersymmetric
gauge theories was shown \cite{egk,egkrs}
to be manifest in string theory. The electric
and magnetic theories provide different
parametrizations of the same quantum moduli
space of vacua. They are related
by smoothly exchanging fivebranes in an
appropriate brane configuration. Many additional
features of the vacuum structure of $N=1$ SUSY
gauge theories were reproduced by studying the
fivebrane configuration \cite{oz,wittt,kapl}.

\end{enumerate}

\noindent
String theory has vastly more degrees of freedom
than field theory but most of these are irrelevant
at low energies. The conventional view in the past
was that to study low energy gauge dynamics, it is
sufficient to retain the light (\ie\ gauge theory)
degrees of freedom and, therefore, string theory can
not shed light on issues having to do with strong
coupling at long distances.

The recent results mentioned above suggest
that while most of the degrees
of freedom of string theory are indeed irrelevant
for understanding low energy dynamics, there is a
sector of the theory that is significantly larger
than the gauge theory in question that should
be kept to understand the low energy structure.
This sector involves degrees of freedom
living on branes and describing their internal
fluctuations and embedding in spacetime.

Brane dynamics seems to provide a new perspective
on gauge theory; the description in terms
of gauge bosons and quarks
appears as an effective low energy
picture that is useful in some region of the moduli
space of vacua. Different descriptions are useful in
different regions of moduli space, and in some regions
the infrared behavior cannot be given a field
theory interpretation at all. The success of the brane
description in reproducing strong coupling
phenomena such as hidden relations between
different gauge theories leads one
to hope that it can serve as a basis
for an alternative formulation of gauge theory.
Such a formulation
might be useful for computing non-vacuum low energy
properties, \eg\ the masses and interactions
of low lying non-BPS states. At present, its
development awaits a better understanding of
the dynamics of fivebranes in string theory.

The set of gauge theories that have been constructed
to date using branes is rather limited. In
four dimensional theories with $N\le2$ SUSY only
the simplest kinds of matter have been studied
(fundamentals and various two index tensors).
It would be very interesting
to study generic chiral $N=1$ SUSY
gauge theories, which in addition to their
phenomenological appeal
have in general rather
rich dynamics.
Some recent work on brane constructions of
chiral gauge theories appears in
\cite{lykken,lykken'}.

One of the most interesting phenomena discovered
in four dimensional $N=1$ SUSY gauge theory
is Seiberg's duality \cite{nati}.
Seiberg's original work on SQCD
has been generalized in gauge
theory in two directions. One class of examples
\cite{ks,ils} involves introducing two index
tensors with polynomial superpotentials.
The duality properties of this class of theories
have been understood using branes
(and we will discuss a chiral example belonging
to this class later). The other class
\cite{pouliot} has the property that one
of the members of a dual pair has an
orthogonal gauge group
with matter in the spinor representation.
It has still not been constructed in
brane theory. It would
be very interesting to construct the theories
of \cite{pouliot} using branes.

The purpose of this paper is to take a modest step
towards describing non-trivial chiral gauge theories
in four dimensions by studying the physics of branes
near orientifolds.
Orientifolds play an essential role in constructing
orthogonal and symplectic gauge theories on branes
\cite{johnson,egkrs},
and allow one to describe theories with interesting
matter content (see for example \cite{ll,lykken'}).
Therefore, it is
likely that understanding brane dynamics near orientifolds
is important for the program outlined above.

The plan of the paper is as follows.
In section 2 we discuss configurations
of parallel orientifolds and D-branes
preserving sixteen supercharges.
By using the relation between
Montonen-Olive duality and S-duality
of type IIB string theory we show that
an orientifold threeplane ($O3$-plane)
with positive Ramond-Ramond (RR) charge
is exchanged under S-duality with
an $O3$-plane with negative charge
with a $D3$-brane embedded in it.
Considering this system on a transverse
torus we briefly discuss the
disconnected components of the moduli space
of $Dp$-branes near an $Op$-plane wrapped
around $T^{p-3}$, in which
$2^{p-3}$ D-branes are stuck at the orientifold.

In section 3 we discuss four dimensional $N=2$ SYM
with orthogonal or symplectic gauge group obtained
by suspending $D4$-branes between $NS5$-branes near
an orientifold plane. We describe the moduli space
of vacua and show that it agrees with gauge theory.
We also give a worldsheet argument for the fact that
an NS fivebrane which intersects an orientifold plane,
splitting it into two disconnected parts, acts as a
domain wall for RR charge. This was previously
observed from the worldvolume gauge theory point of
view in \cite{johnson}.

In section 4 we study a configuration describing a chiral
$N=1$ SUSY four dimensional gauge theory. We discuss
some of its deformations and show that
the brane analysis reproduces the predictions of
\cite{ils} regarding duality in this system.

\section{ Systems With Sixteen Supercharges}

Consider a configuration of $\nc$ $D3$-branes
and an $O3$-plane all
stretched in $(x^0, x^1, x^2, x^3)$. The threebrane
worldvolume gauge theory has $N=4$ SUSY. The
orientifold projection breaks the usual unitary
gauge group on $\nc$ parallel $Dp$-branes
to a symplectic or orthogonal one, depending
on the RR charge of the orientifold.
For an orientifold of charge $Q_{O3}=+Q_{D3}/2$
($Q_{O3}=-Q_{D3}/2$) the gauge group is $G=Sp(\nc/2)$
($G=SO(\nc)$). In the former case $\nc$ must
be even while in the latter it can be odd.
For odd $\nc$ there is a single $D3$-brane
without a mirror image stuck at the orientifold
plane. It is easy to see that such a brane
survives the orientifold projection only for
$Q_{O3}=-Q_{D3}/2$.

The gauge coupling of the theory on the threebrane
$g$ is related to the IIB string coupling $\gs$
via $g^2=\gs$. If we send the string length
$l_s\to 0$ holding $g$ fixed, the theory on the
threebranes decouples from gravity and massive
string modes, and the four dimensional dynamics
becomes that of $N=4$ SYM with gauge group
$G$ at all energy scales.

The Coulomb branch of the theory is
parametrized by the locations of the $D3$-branes
in the transverse space
labeled\footnote{In the limit $l_s\to0$
one must hold the energy scale
$\Phi^i=x^i/l_s^2$ ($i=4,\cdots,9$) fixed.} by
$(x^4,x^5, \cdots, x^9)$. Since the
$D3$-branes can only leave the
orientifold plane in pairs, the
dimension of the Coulomb branch is
$6\times [\nc/2]$ as expected from gauge theory.
Generically in the Coulomb branch the
gauge symmetry is broken to $U(1)^{[\nc/2]}$,
and all the charged gauge bosons and dyons,
which correspond to $(p,q)$
strings stretched between different
threebranes,
are massive. Singularities in moduli space
correspond to points with enhanced unbroken
gauge symmetry, at which charged gauge
bosons and dyons go to zero mass. The
most singular point in the moduli space
is the origin, which
corresponds to a non-trivial CFT parametrized
by the exactly marginal gauge coupling $g$.

One application of this construction
is to the study of the moduli space of
monopoles in broken $SO(\nc)$ or $Sp(\nc/2)$
gauge theories \cite{diac}.
Monopoles in broken $SO/Sp$ gauge theory are described
by $D$-strings stretched between different
$D3$-branes. Consider for example the rank one case
$\nc=2$. For positive  orientifold charge the gauge
group is $Sp(1)\simeq SU(2)$ and the moduli space of
$k$ $SU(2)$ monopoles
can be studied by analyzing the worldvolume theory
of $k$ $D$-strings connecting the single ``physical''
$D3$-brane to its mirror image.
For negative orientifold charge the gauge group is
$SO(2)\simeq U(1)$ and one does not expect non-singular
monopoles to exist. This means that $D$-strings cannot
connect the single physical $D3$-brane to its mirror
image. This is related by S-duality to the fact
that for negative orientifold charge the
ground states of fundamental strings
stretched between the $D3$-brane and its image are
projected out by the orientifold projection.

The identification of Montonen-Olive
duality with S-duallity of type IIB
string theory can be used
to learn about strong coupling
properties of orientifolds. Recall that in gauge
theory, electric-magnetic duality acts on the
gauge coupling as strong-weak coupling
duality $g\to 1/g$. It also takes the gauge
algebra to the dual algebra; in particular
$G=SO(2r)$ is self-dual, while
$SO(2r+1)$ and $Sp(r)$ transform to
each other under Montonen-Olive duality.

The $D3$-brane is self-dual under S-duality,
and so is the four-form gauge field
that it couples to.
For agreement with $N=4$ supersymmetric $SO(2r)$
gauge theory, the $O3$-plane with negative charge
must be self-dual as well. To study the
non-simply laced case
consider \eg\ a weakly coupled $SO(2r+1)$ gauge theory.
The orientifold charge is $-Q_{D3}/2$; the $6r$ dimensional
Coulomb branch corresponds to removing $r$
pairs of threebranes from the orientifold
plane. A single threebrane which does not have a
mirror remains stuck at the orientifold.
When the gauge coupling becomes large there are
two ways of thinking about the system. We can
either continue thinking about it as a strongly
coupled $SO(2r+1)$ gauge theory, or
relate it to a weakly coupled $Sp(r)$
gauge theory by performing
a strong-weak coupling S-duality transformation.
The latter is described
by an orientifold with charge $+Q_{D3}/2$.

Thus, Montonen-Olive duality of gauge theory teaches
us that a ``bound state'' of an $O3$-plane with negative
RR charge and a single $D3$-brane embedded in it
(a configuration with RR charge $(-1/2+1)Q_{D3}$)
transforms under S-duality of type IIB string theory
into an $O3$-plane with Ramond charge $+Q_{D3}/2$.
This is consistent with the fact that RR four-form charge
cannot change under S-duality, since the corresponding
gauge field is self-dual.

For higher dimensional orientifolds, similar
considerations lead to a strong-weak coupling
relation between the following two apparently
different configurations. One has an $Op$-plane
with positive charge wrapped around a $p-3$
dimensional torus $T^{p-3}$. The other is
a certain disconnected component of the moduli
space of $2^{p-3}$ $Dp$-branes in the presence
of a negatively charged orientifold plane,
$Op_-$ (all objects wrapped around a dual
torus $\widehat T^{p-3}$),
in which the $2^{p-3}$ branes are
stuck at the orientifold.

The relation is obtained by considering
an $Op$-plane with positive RR charge
wrapped around
$T^{p-3}$. Clearly, this configuration
has no moduli and this must be the case
for any other configuration related to
it by U-duality.
Performing a T-duality transformation
which inverts the volume of the $(p-3)$
torus we find $2^{p-3}$ $O3$-planes
with positive charge
distributed at equal distances on
the dual torus $\tilde T^{p-3}$.
A further S-duality turns them
into $2^{p-3}$ negative charge
orientifold planes, each of which
has a $D3$-brane embedded in it.
A final T-duality on the $(p-3)$
torus gives rise to an $Op$-plane
with negative charge wrapped around
a torus $\widehat T^{p-3}$ and
$2^{p-3}$ $Dp$-branes embedded
in it. Since the original system we started
with (the positive charge $Op$-plane)
did not have any moduli, this must
be the case for the final configuration
as well. In particular, the $2^{p-3}$
D-branes are stuck at the orientifold,
without the ability to move.
Similar disconnected components of the
moduli space of D-branes near orientifold
planes have been recently discussed in
\cite{edtor}.

\section{Systems With Eight Supercharges}

Four dimensional $N=2$ SUSY gauge theory
with gauge group $G=U(\nc)$ and $\nf$
hypermultiplets in the fundamental
representation of the gauge group
can be studied \cite{hanany,egk}
by suspending $\nc$
$D4$-branes with worldvolume
$(x^0, x^1, x^2, x^3, x^6)$
between two $NS5$-branes which are
extended in
$(x^0, x^1, x^2, x^3, x^4, x^5)$
and located at different values of $x^6$.
The fourbranes are finite in $x^6$,
stretching between the two $NS5$-branes;
thus they describe four dimensional
physics on their worldvolume. The matter
hypermultiplets arise from $\nf$
$D6$-branes extended in
$(x^0, x^1, x^2, x^3, x^7, x^8, x^9)$ and
placed between the two $NS5$-branes in
$x^6$ (see
\cite{gk} for a more detailed discussion).

To describe orthogonal and symplectic
gauge groups one can add to the brane
configuration an $O4$-plane parallel
to the $D4$-branes, or an $O6$-plane
parallel to the $D6$-branes
\cite{johnson,egkrs}. Both
cases are going to be useful below
when we describe the chiral $N=1$
configuration. We will next discuss
the better understood case of an
$O6$-plane, confining our discussion
of the $O4$-plane to a few comments.
A more detailed discussion appears
in \cite{gk}.

In the presence of an $O6$-plane
we would like to stretch $\nc$
fourbranes from an $NS5$-brane
to its mirror image. The first
question that we have to address
is whether it is possible to
stretch fourbranes this way
without breaking SUSY. For example,
in the previous section\footnote{Where
we discussed explicitly the case $p=1$;
clearly the result is $p$ independent.}
we saw that
it is impossible to stretch a BPS saturated
$Dp$-brane between a $D(p+2)$-brane
and its mirror image with respect to
an $O(p+2)$-plane with negative charge.
One might worry that the same is true
for $D4$-branes stretched between
$NS$-branes.

In fact, it turns out that the relevant
$D4$-branes are indeed BPS saturated
\cite{gk}.
To show that, one maps (using U-duality)
the fourbrane connecting an $NS5$-brane
to its image, to a fundamental string
connecting two $D5$-branes in the presence
of an orientifold nineplane, and uses known
properties of D-branes.

We will next describe the classical gauge
theories arising for the two choices
of the sign of the orientifold charge,
starting with
the case of positive $O6$ charge, which
leads to an
orthogonal projection on the $D4$-branes.
The case of
$O6_-$, which leads to a symplectic
gauge group, will be considered later.

The gauge group on $\nc$ $D4$-branes
connecting an $NS5$-brane to its mirror
image with respect to an $O6_+$ plane
is $SO(\nc)$.
$\nf$ $D6$-branes parallel to the $O6$-plane
located
between the $NS5$-brane and the orientifold
give
$\nf$ hypermultiplets in the fundamental
(${\bf \nc}$) representation of $SO(\nc)$,
arising as usual from $4-6$ strings.
In $N=1$ SUSY notation there are
$2\nf$ chiral multiplets $Q^i$, $i=1,\cdots, 2\nf$
which are paired to make $\nf$ hypermultiplets.
The global flavor symmetry of this gauge theory
is $Sp(\nf)$, in agreement with the
projection imposed by the positive charge
$O6$-plane on the $D6$-branes.

The Coulomb branch
of the $N=2$ SUSY gauge theory is parametrized
by the locations of the $D4$-branes along the
fivebrane, in the $(x^4, x^5)$ plane. Entering the
Coulomb branch involves removing the ends of the
fourbranes from the orientifold plane (which is
located at a particular point in the $(x^4, x^5)$
plane). Since the fourbranes can only leave the
orientifold plane in pairs, the dimension of the
Coulomb branch is $[\nc/2]$, in agreement with
the gauge theory description.

The different Higgs branches of the gauge theory
are parametrized by all possible breakings of
fourbranes on sixbranes. As for the unitary
case there are many different branches;
as a check that we get the right structure,
consider the fully Higgsed branch
which exists when the number of
flavors is sufficiently large. {}From gauge
theory we expect its dimension to be
$2\nc\nf-\nc(\nc-1)$.

The brane analysis gives
\beq
{\rm dim} \MM_H=\sum_{i=1}^{\nc}
2(\nf+1-i)=2\nf\nc-\nc(\nc-1)=[2\nf\nc-\nc(\nc+1)]+2\nc
\label{OSG1}
\eeq
The term in the square brackets is the number of
moduli corresponding to segments that do
not touch the orientifold, and the additional $2\nc$
is the number of moduli coming
from the segments of the fourbranes connecting
the $D6$-brane closest to the orientifold to its
mirror image. These segments transform to
themselves under the orientifold projection
and thus are dynamical for positive orientifold
charge.

The $2\nc$ moduli coming from fourbranes
connecting a $D6$-brane to its image have
a natural interpretation in the theory on
the $D6$-branes. At low energies this is
an $Sp(1)$ gauge theory with sixteen
supercharges, and the $D4$-branes stretched
between the $D6$ and its mirror can be thought
of as $Sp(1)$ monopoles, as in the previous section.
{}From this point of view the above
$2\nc$ moduli parametrize the space of $\nc$
$Sp(1)$ monopoles.

Thus, the total dimension of moduli space agrees
with the gauge theory result. It is easy
to similarly check the agreement with gauge theory
of the maximally Higgsed branch for small
$\nf$, as well as the structure of
the mixed Higgs-Coulomb phases. The analysis
of the vacuum structure of the corresponding
gauge theories appears in \cite{aps}.

For negative charge of the $O6$-plane, the configuration
discussed above describes an $Sp(\nc/2)$ gauge theory
with $\nf$ hypermultiplets in the fundamental
$({\bf \nc})$ representation. Qualitatively,
most of the analysis is the
same as above, but the results are clearly
somewhat different. For example, the dimension of the
fully Higgsed branch is in this case $2\nf\nc-\nc(
\nc+1)$, smaller by $2\nc$ than the $SO$ case discussed
above.

{}From the point of view of the brane construction
the Higgs branch is different because it is no longer
possible to connect a $D6$-brane to its mirror image
by a $D4$-brane. Such fourbranes transform to themselves
under the orientifold projection, and are
projected out when the $O6$-plane has negative charge.
This is also clear from the point of view of interpreting
these fourbranes as magnetic monopoles in the sixbrane
theory. In this case the theory on the $D6$-brane
adjacent to the orientifold and its image
has gauge group $SO(2)$, and there are no non-singular
monopoles.

Therefore, the breaking of the $D4$-branes
on $D6$-branes near the orientifold is modified.
We have to stop the
pattern (\ref{OSG1}) when we get to the last {\em two}
$D6$-branes before the orientifold, and there we must
perform the breaking in such a way that no
fourbrane connects a $D6$-brane to its mirror image.
It is not difficult to see \cite{gk} that
compared to (\ref{OSG1}) we lose $2\nc$ moduli.
Overall, the brane
Higgs branch is $2\nf\nc-\nc(\nc+1)$ dimensional,
in agreement with the gauge theory analysis. One can
again check that the full classical phase structure of the
$Sp(\nc/2)$ gauge theory \cite{aps} is similarly reproduced.

Another way to study
four dimensional $N=2$ SUSY
gauge theories with symplectic
and orthogonal gauge groups is by using
branes and orientifold fourplanes.
It is interesting and in some ways
more mysterious than the construction
using $O6$-planes described above
(see \cite{gk} for a more detailed discussion).

Consider, for example, a pair of $NS5$-branes
with worldvolume $(x^0,x^1,\cdots, x^5)$
stuck on an orientifold
fourplane stretched in $(x^0,x^1,x^2,x^3,
x^6)$. The two $NS5$-branes are separated
by a distance $L_6$ (in the $x^6$ direction)
along the orientifold.
We can now stretch $\nc$ $D4$-branes parallel
to the $O4$-plane between the two $NS5$-branes,
and place $2\nf$ $D6$-branes with worldvolume
$(x^0, x^1, x^2, x^3, x^7, x^8, x^9)$ between the
$NS5$-branes as before.
Alternatively, we can replace the $D6$-branes
by $2\nf$ semi-infinite $D4$-branes attached to the
$NS5$-branes and extending (say) to $x^6=\pm \infty$.

Each of the $NS5$-branes divides the orientifold
into two disconnected parts and acts as a domain
wall for orientifold charge. If the charge on one
side of the fivebrane is $+1$, on the other it is
$-1$ and vice versa \cite{johnson}.
We will next motivate this interesting effect
by comparing the brane picture to the worldvolume
gauge theory, as well as by a direct worldsheet
analysis.

To see that the orientifold charge flip
is required by gauge theory it is useful
to represent fundamental hypermultiplets
by semi-infinite fourbranes.
If the charge of the
segment of the $O4$-plane trapped between
the $NS5$-branes is positive, the gauge
group on $\nc$ mirror pairs of
$D4$-branes stretched between the $NS5$-branes
is $Sp(\nc)$. $2\nf$ semi-infinite $D4$-branes
extending to $x^6=\infty$ (say) give rise to
$\nf$ hypermultiplets in the fundamental
representation of $Sp(\nc)$.
It is well known (see \eg\ \cite{aps})
that the global symmetry in gauge theory in this
situation is $SO(2\nf)$. In brane theory, the
global symmetry arises from the gauge symmetry
of the $2\nf$ semi-infinite fourbranes. To get the
right global symmetry,
the charge of the segment of the orientifold
to the right of the $NS5$-branes must be negative.

An alternative gauge theory explanation of the
charge flip involves a six dimensional version
of the above brane construction that will be useful
in the next section. The four dimensional $N=2$
SUSY gauge theory in question can be thought of
as a dimensional reduction of a six dimensional
$N=1$ SUSY theory. To obtain the six dimensional
theory using branes, we can compactify $(x^4, x^5)$,
T-dualize the configuration, and then decompactify
the dual $(x^4, x^5)$. Using the standard
action of T-duality on branes \cite{gk}, one finds
that
the $D4$-branes and $O4$-plane become
$D6$-branes and an $O6$-plane stretched in
$(x^0,x^1,x^2,x^3,x^4, x^5, x^6)$; the $NS5$-branes
are invariant.
Such brane configurations realizing six dimensional
theories were studied in \cite{Brunner}.

The $5+1$ dimensional $Sp(\nc)$ gauge
theory with $\nf$ hypermultiplets is anomalous
for generic $\nf$. It is anomaly free
for $\nf=2\nc+8$.
Anomaly freedom in six dimensions
manifests itself in the brane construction as RR charge
conservation on the $NS5$-branes. The $D6$-branes
ending on them carry net RR charge that has nowhere
to escape and thus has to be cancelled.
We can split the $2\nc+8$ flavors symmetrically
between the two $NS5$-branes, by attaching
$2\nc+8$ semi-infinite $D6$-branes to the right
of the rightmost $NS5$-brane and the other
$2\nc+8$ to the left of the leftmost $NS5$-brane.
Naively, there appears to be a deficit of eight units
of charge on each $NS5$-brane. It is cancelled
by the jump of the orientifold charge from $+4$
to $-4$ described above\footnote{In general,
the charge of an $Op$-plane cut into two parts
by an $NS5$-brane jumps from $+2^{p-4}$
to $-2^{p-4}$.}.

To understand the jump of the charge of
an $Op$-plane cut in two by an $NS5$-brane
directly in worldsheet terms, recall that
the RR charge of the orientifold can be
computed by evaluating the one point function
of the vertex operator of the
closed string RR $(p+1)$-form
gauge field to which the
$Dp$-brane couples,
on the leading non-orientable
Riemann surface, $RP^2$. This diagram
is non zero and is localized at the orientifold.
The charge of the orientifold to the left
of the $NS5$-brane is given by a worldsheet
path integral over an $RP^2$ surface embedded
in spacetime
with $x^6$ large and negative. We may refer
to this surface as $RP^2_L$.
The charge of the segment to the right
of the $NS5$-brane is given by a path integral
over a surface with $x^6$ large and positive,
which we will refer to as $RP^2_R$.

The path integral contains the term
$\exp i(\int_{RP^2}B)$ where $B$ is the
NS-NS sector Kalb-Ramond two-form gauge field
under which the $NS5$-brane is magnetically
charged.
The ratio of the path integrals
corresponding to the one point function
of the RR $(p+1)$-form gauge field
far to the left and to the right of
the $NS5$-brane (which gives the
ratio of the corresponding RR charges
$Q_L/Q_R$)
can thus be written by using
Stokes' theorem as
\beq
Q_L/Q_R=\exp i\int_{C_3}H
\label{chrg}
\eeq
where $H=dB$,
and $C_3$ is a three dimensional surface
whose boundaries are $RP^2_L$ and
$RP^2_R$. It is important that $C_3$
necessarily encloses the $NS5$-brane
and thus the integral receives a contribution
which is due to
the magnetic coupling of the $NS5$-brane
to $B$.
If we replaced $RP^2$ by a two-sphere
in the above analysis, the integral
(\ref{chrg}) would give $2\pi$, the total flux
of the $NS5$-brane (by the Dirac quantization
condition). Since the boundary is $RP^2$
the result is $\pi$, half of the total
flux\footnote{No inconsistency with the
Dirac quantization condition is induced
by this, since there are no surfaces
$C_3$ with boundaries $RP^2_L$ and
$RP^2_R$ which do not enclose the
$NS5$-brane.}.
Substituting this into (\ref{chrg})
we see that the charges of the left and right
portions of the orientifolds are opposite,
$Q_L=-Q_R$.

The construction of four dimensional
$N=2$ SYM using branes near an $O4$-plane
can be used to study the moduli space
of vacua of the theory. Some of the
details of this analysis appear in
\cite{gk}.

\section{Systems With Four Supercharges}

In the previous section we discussed configurations
of $NS5$, $D4$ and $D6$-branes in the presence of
orientifold planes, which preserve eight supercharges
and are useful for describing four dimensional $N=2$
SUSY gauge theories. To describe $N=1$ SYM we would
like to
break four supercharges by changing the orientation
of some of the branes in the configuration
\cite{egk}. Defining
the complex variables $v$, $w$ by:
\bea
v=&x^4+ix^5\nonumber\\
w=&x^8+ix^9
\label{D4N12}
\eea
one can check \cite{bar} that relative
complex rotations of the different branes
and orientifolds in the $(v,w)$ plane,
\beq
\left(
\begin{array}{c}
v\\
w\end{array}\right)
\longrightarrow
\left(\begin{array}{cc}
\cos\theta&\sin\theta\\
-\sin\theta&\cos\theta
\end{array}\right)
\left(\begin{array}{c}
v\\
w\end{array}\right)
\label{BST38}
\eeq
do the job. For generic rotation
angles $\{\theta_i\}$ the configuration
preserves four supercharges.
For particular values of
$\{\theta_i\}$ the SUSY is
enhanced to eight supercharges,
and one recovers the configurations
of section 3.

As an example, the $NS5$-brane used
in section 3 was located at a particular
value of $w$ and stretched in $v$.
Rotating it
by an angle $\theta$ (\ref{BST38}) we find
a fivebrane which we may call
the ``$NS_{\th}$ fivebrane.'' It is
located at $w=v\tan\th$ and stretched
in the orthogonal direction,
$v_\th=v\cos\th+w\sin\th$. Two
particularly useful special cases are
$\th=0, \pi/2$. The first corresponds
to the original $NS5$-brane of section 3.
The second is the \nsp-brane of \cite{egk},
which is stretched in $w$ and located at a particular
value of $v$.

Similarly, one can rotate each of the $D6$-branes
of section 3 separately as in (\ref{BST38}).
This leads to a large variety
of $N=1$ SUSY models which we will not discuss
here (see \cite{gk} for more details).
Instead we will analyze below a particular
example -- a chiral $N=1$ model that illustrates some
of the issues discussed in the previous sections.

\subsection{A Chiral Brane Configuration}

Below we will
consider brane configurations
in which an \nsp-brane without a mirror
image is embedded in an $O6$-plane
extending, as in section 3, in
$(x^0,x^1,x^2,x^3,x^7,x^8,x^9)$.
One can achieve this situation
by embedding
a pair of \nsp-branes
in the $O6$-plane, separating them
along the orientifold (in $x^7$)
and taking one of them to infinity.
The remaining \nsp-brane is free to
move inside the $O6$-plane, along
the $x^7$ axis. In the absence of
other branes nothing depends on its
location; after we add more features
it will become an important parameter
on which low energy physics depends
sensitively.

The \nsp-brane,
located (say) at $x^7=0$, divides
the $O6$-plane into two disconnected
regions, corresponding to positive
and negative $x^7$. As we saw
in section 3, in this situation
the RR charge of the orientifold jumps,
from $+4$ to $-4$, as we cross the
\nsp-brane. The part of the orientifold
with negative charge (which we will
take to correspond to $x^7<0$) has furthermore
eight semi-infinite $D6$-branes
embedded in it.
As discussed in section 3, the
presence of these $D6$-branes
is required for charge conservation
or, equivalently, vanishing of the
six dimensional anomaly.

In addition to the eight semi-infinite
$D6$-branes, we can place on the
orientifold any number of parallel
infinite $D6$-branes extending all the
way from $x^7=-\infty$ to $x^7=\infty$.
We will denote the number of such
$D6$-branes by $2\nf$.

Then, an $NS_{\th}$ fivebrane located
at a distance $L_6$ in the $x^6$ direction
from the \nsp-brane, but at the same value
of $x^7$, is connected to the
\nsp-brane by $\nc$ $D4$-branes stretched
in $x^6$, as in section 3. We will see later
that $\nc$ must be even for consistency.
The mirror image
of the $NS_{\th}$ fivebrane, which is an
$NS_{-\th}$ fivebrane, is necessarily also
connected to the \nsp-brane.

We can also place any number of $D6$-branes
oriented at arbitrary angles $\theta_i$
(\ref{BST38}) between the $NS_\th$ fivebrane
and the orientifold (in $x^6$). We will
mainly discuss the case where such branes are
absent, but it is easy to incorporate them,
following \cite{aharony,gk}.

As is by now standard \cite{gk},
at low energies and weak string
coupling\footnote{More precisely,
in the limit $l_s,\gs, L_6\to0$,
with the gauge coupling
$g^2=\gs l_s/L_6$ held fixed.}
the brane configuration describes
a four dimensional gauge theory.
Our next task is to identify the
particular gauge theory that arises.
In the next subsections we address
this problem. Before studying
the general case we describe the
structure for $\theta=0$ (when
the external $NS_{\pm \th}$ fivebranes
are $NS5$-branes), and $\theta=\pi/2$
(when they are \nsp-branes).
We also show that the field theory
duality of \cite{ils} for the
resulting gauge theory is reproduced
by the brane analysis \cite{egk}.

\subsection{The Case $\th=0$}

The gauge theory corresponding to the brane
configuration described above can be pieced
together as follows. Note that
the configuration depends on a real parameter $r$
corresponding to the relative position in $x^7$
of the $NS5$-brane (and its mirror image) and
the \nsp-brane embedded in the $O6$-plane. In the
original configuration
$r$ vanishes, but it is useful to study the
physics as a function of it.

For positive $r$,
the $D4$-branes connecting the $NS5$-brane to the
\nsp-brane must reconnect to stretch between the
$NS5$-brane and its mirror image.
At low energies we
can ignore the \nsp-brane and the region
$x^7<r$, and find the brane configuration
analyzed in section 3 -- an $NS5$-brane
connected to its image with respect to
an $O6_+$ plane by
$\nc$ $D4$-branes, in the
presence of $2\nf$ $D6$-branes
parallel to the orientifold.
The low energy theory has in this case
an accidental $N=2$ SUSY; the gauge
group is $G=SO(\nc)$
and the matter corresponds to
$\nf$ hypermultiplets ($2\nf$ chiral multiplets)
in the fundamental representation.
Denoting the adjoint chiral multiplet
in the $SO(\nc)$ vectormultiplet by
$A$ and the fundamental chiral multiplets
by $\tilde Q_a$, the $N=2$ superpotential is
\beq
W=\tilde QA \tilde Q
\label{ntwoso}
\eeq
The case of negative $r$ is similar,
except now the charge of the $O6$-plane
felt by the $D4$-branes intersecting it
is negative, and there are $2\nf+8$ $D6$-branes
embedded in it. The corresponding gauge theory
is an $N=2$ SUSY $Sp(\nc/2)$ gauge theory
(this makes it clear that $\nc$ must be even)
with $\nf+4$ hypermultiplets ($2\nf+8$ chiral
multiplets $Q^i$) in the
fundamental representation.
The $N=2$ superpotential couples the
adjoint chiral multiplet $\tilde S$ to the
fundamentals
\beq
W=Q \tilde SQ
\label{ntwosp}
\eeq
What can we say about the theory with $r=0$?
In the absence of the orientifold, the gauge
group would be $U(\nc)\times
U(\nc)$ \cite{gk}. The orientifold projection
preserves the diagonal $U(\nc)$. This is consistent
with the low energy gauge groups for
$r\not=0$, both of which are subgroups
of $U(\nc)$. $r$ is interpreted, as in
other similar situations in brane theory
\cite{gk}, as a FI D-term for the $U(1)$
factor in $U(\nc)$.

As $r\to 0$ we expect all the light fields
seen for either sign of $r$ to become
massless. Thus, the $U(\nc)$ gauge group
at the origin couples to an antisymmetric
tensor $A$, a symmetric tensor $\tilde S$,
$2\nf+8$ quarks $Q$ in the fundamental
representation, and $2\nf$ quarks $\tilde Q$ in the
antifundamental representation. The superpotential
is a combination of (\ref{ntwoso}, \ref{ntwosp}):
\beq
\label{eq:sup1}
W=Q\tilde{S}Q+\tilde{Q}A\tilde{Q}
\eeq
The transformation
properties of the quarks
can be understood following
\cite{hb,aharony,hz} who argued that
fundamental chiral multiplets
of the gauge group come from
$4-6$ strings connecting the $D4$-branes
to $D6$-branes ending on an \nsp-brane
from below (in $x^7$), while antifundamentals
arise from $D6$-branes ending on the
\nsp-brane from above. In our case there
are therefore $2\nf+8$ fundamentals $Q^i$, and
$2\nf$ antifundamentals $\tilde Q_a$.
The global symmetry
of the system is determined by the gauge symmetry
on the $D6$-branes, $Sp(\nf)\times SO(2\nf+8)$.
The superpotential (\ref{eq:sup1}) is the unique
one consistent with this symmetry.

Note that the theory is chiral and potentially
anomalous as there are eight more fundamental
than antifundamental chiral multiplets.
The superpotential
(\ref{eq:sup1}) implies that
the symmetric tensor $\tilde S$ is in fact
a symmetric bar (\ie\ a symmetric
tensor with two antifundamental indices).
Thus the total anomaly
$(2\nf+8)-2\nf+(\nc-4)-(\nc+4)$ vanishes,
as one would expect for a consistent
vacuum of string theory.

An interesting feature of the spectrum
of light fields is the fact that there are
$2\nf$ fundamental superfields $Q^i$ and
$2\nf$ antifundamentals $\tilde Q_a$ that can be
given a mass. Recall \cite{gk} that masses correspond
to displacements of the $D6$-branes relative to
the $D4$-branes (and thus to the orientifold as well)
in the $(x^4, x^5)$ directions.  Since
one can only remove $D6$-branes from the
orientifold in pairs, only $\nf$ independent
mass parameters seem to be visible, which
would appear to indicate that there are only
$\nf$ flavors.

To see why there are in fact $2\nf$ flavors
it is convenient to remove the $D6$-branes
from the orientifold in the $x^6$ direction.
Positions of $D6$-branes along the $x^6$
axis are often irrelevant parameters in the low
energy limit of brane theory. However, it is
well known that precisely in the case relevant here,
the relative position of an \nsp-brane and
a $D6$-brane is important for the
infrared dynamics. When
a $D6$-brane crosses an \nsp-brane
(more generally, when parallel
D and NS branes cross)
the theory loses or gains a massless
flavor \cite{gk}.

In our case when a $D6$-brane and its image are
displaced in $x^6$ from the orientifold
(but remain at the origin in $x^4$, $x^5$)
two of the four degrees of freedom arising
from the corresponding $4-6$ strings gain
a mass, while two remain massless. The
massive states correspond to strings
stretched between the $D6$-branes and
$D4$-branes on the opposite side of the orientifold.
These strings have a finite length and hence
describe massive states. The massless degrees
of freedom correspond to short $4-6$ strings
connecting the sixbranes to fourbranes
on the same side of the orientifold plane.
Of course, as the $D6$-brane approaches the
orientifold and the \nsp-brane
embedded in it (in $x^6$), the previously
massive matter becomes light. When the
sixbrane and orientifold coincide the number
of massless states jumps by a factor of two.
This is why the theory has $2\nf$ fundamental --
antifundamental pairs.

As a further check on the identification of the
brane configuration and the chiral gauge theory
one can analyze the moduli space of vacua
as a function of various parameters one can add to
the Lagrangian. An example is the FI D-term, which
we have
already discussed from the
brane point of view.
In the gauge theory, adding to the Lagrangian
a FI D-term for the
$U(1)$ vectormultiplet ${\rm Tr} V$,
\beq
\LL_D=r\int d^4\theta\, {\rm Tr} V
\label{rd}
\eeq
modifies the
D flatness vacuum conditions:
\beq
A A^\dagger -\tilde S \tilde S^\dagger
+Q Q^\dagger -\tilde Q \tilde Q^\dagger
=-r
\label{dtermflat}
\eeq
Setting the quarks $Q$, $\tilde Q$
to zero we see that
when $r$ is positive, $S$ gets an expectation
value which breaks $U(\nc)\to SO(\nc)$.
Due to the superpotential
(\ref{eq:sup1}) the
$2\nf+8$ chiral multiplets $Q^i$
as well as $\tilde{S}$ become
massive and one if left with the
$N=2$ spectrum and interactions
for gauge group $SO(\nc)$,
(\ref{ntwoso}), with the antisymmetric
tensor $A$ playing the role of the adjoint
of $SO(\nc)$. All of this is
easily read off the brane configuration.
In particular, the fact that the $2\nf+8$
quarks $Q^i$ are massive is due to the fact
that the corresponding $4-6$ strings
have finite length (proportional to $r$).

Similarly, for negative $r$ (\ref{dtermflat})
implies that $A$ gets an expectation value,
breaking $U(\nc)$ to $Sp(\nc/2)$. The quarks
$\tilde Q$ get a mass and we end up with
an $N=2$ gauge theory with $G=Sp(\nc/2)$
and $2N_f+8$ light quarks.

The Coulomb, Higgs
and mixed branches of moduli space are
realized in a similar way to that described
in section 3 and in \cite{gk}.
The Coulomb branch is obtained by removing
the $D4$-branes from the orientifold plane
in pairs along the $NS5$-brane and its mirror
image (\ie\ in the $v$-plane). This gives
rise to an $\nc/2$ dimensional moduli
space which in gauge theory corresponds to
giving expectation values to the fields $A$,
$\tilde S$. Since when we enter the Coulomb branch
the fourbranes are already reconnected and
stretch between the $NS5$-brane and its image,
it doesn't matter whether the FI D-term $r$
is turned on or not. Thus the structure
of the Coulomb branch is very similar to
that of the corresponding $N=2$ SUSY theory,
and we will not discuss it in any detail
here\footnote{Except to note that the brane
configuration in question can be used to show that
the Coulomb branches of $SO(\nc)$ gauge theory
with $\nf$ hypermultiplets, and $Sp(\nc/2)$
gauge theory with $\nf+4$ hypermultiplets four of
which are massless, are
identical. This can be checked directly
by comparing the appropriate Seiberg-Witten curves.}.
The Higgs branch of the $N=2$ SUSY theories was
described using branes in section 3. There is
a jump in the dimension of the Higgs branch
as we pass through $r=0$ -- the Higgs branches
of the symplectic and orthogonal theories
are different. This too is clearly seen in the
brane construction.

\subsection{The Case $\th=\frac{\pi}{2}$}

In this case the external fivebrane and its mirror
image are \nsp-branes. If the orientifold
was absent, the configuration would correspond
to a $U(\nc)\times U(\nc)$ gauge theory with
two chiral multiplets $\Phi_1$, $\Phi_2$
transforming
as $(\nc^2,1)$ and $(1,\nc^2)$, respectively.
$\Phi_1$ and $\Phi_2$ parametrize separate motions
of the two groups of $\nc$ fourbranes along
the fivebranes (\ie\ in the $w$ direction).
The orientifold again preserves one of the two
$U(\nc)$ factors, and the corresponding
adjoint field $\Phi$.

The Coulomb branch
is now $\nc$ dimensional. It is labeled
by locations in the $w$ plane of the
$\nc$ $D4$-branes stretched between the
\nsp-brane outside the orientifold and the
one inside it. Of course, the $\nc$ mirror
$D4$-branes on the other side of the orientifold
(in $x^6$) follow the same motion.
Generically along the Coulomb branch the
gauge group is broken to $U(1)^{\nc}$.

The corresponding gauge theory has in addition
to the matter content discussed in the
previous subsection the adjoint field
$\Phi$. The classical superpotential
generalizing (\ref{eq:sup1}) is:
\beq
\label{eq:sup3}
W=Q\tilde{S}Q+\tilde QA\tilde{Q}+\Phi A\tilde{S}
\eeq
Because of the last term in $W$, generically
in the Coulomb branch $A$, $\tilde S$ are massive
(except for the diagonal components of $\tilde S$).
In the brane construction we saw that off-diagonal
components of $A$
and $\tilde S$ are described by strings
connecting different fourbranes on
opposite sides of the orientifold (while
$\Phi$ is described by strings connecting
different fourbranes on the same
side of the orientifold). When
the $D4$-branes are separated in $w$,
such strings become long and thus
the corresponding components of
$A$, $\tilde S$ become massive.

The moduli space associated with
giving expectation values to
$A$, $\tilde S$ is absent here.
In the brane language this is
because the fourbranes
are not free to move in $v$; in
gauge theory it is due to the superpotential
(\ref{eq:sup3}), one of whose equations
of motion is
\beq
\label{eq:motion1}
\tilde{S}A=0
\eeq
which together with the D flatness condition
(\ref{dtermflat})
sets $A$, $\tilde S$ to zero.

As a check on the gauge theory we can again
study the D-term perturbation (\ref{rd}) corresponding
to relative displacement of the \nsp-branes. For
positive $r$ we now find an $SO(\nc)$ gauge theory
with $2\nf$ fundamental chiral multiplets, a symmetric
tensor and vanishing superpotential. This can be understood
by analyzing the D-flatness conditions (\ref{dtermflat})
in the presence of the superpotential (\ref{eq:sup3}).
As before, the symmetric tensor $\tilde S$ gets
an expectation value, which for unbroken $SO(\nc)$
must be proportional to the identity matrix.
The last term in the superpotential (\ref{eq:sup3})
then gives rise to the mass term $W\sim \Phi A$.
Since $A$ is antisymmetric, this term gives a mass
to the antisymmetric part of $\Phi$ (as well as to
$A$). The symmetric part of $\Phi$ becomes the
symmetric tensor mentioned above. Clearly, it
does not couple to the $2\nf$ fundamental chiral multiplets.
In the brane description the fact that fluctuations
of the fourbranes in $w$ are described by a symmetric
tensor is a direct consequence of the action
of the orientifold projection \cite{gimon}.

The resulting $SO(\nc)$ gauge theory has an
$\nc$ dimensional moduli space corresponding to
giving an expectation value to the symmetric tensor.
Generically in this moduli space the gauge group is
completely broken\footnote{As a check,
the symmetric tensor has $\nc(\nc+1)/2$ components
out of which $\nc(\nc-1)/2$ are eaten up by the
Higgs mechanism. The remaining $\nc$ massless
fields parametrize the moduli space.}.
In the brane description this
moduli space corresponds to separating the $\nc$
$D4$-branes along the orientifold plane (in $w$).
Generically in the moduli space no two fourbranes
are at the same $w$ and the gauge symmetry is completely
broken. The fact that the symmetric tensor does not
couple to the fundamentals is reflected in the brane
construction by the property that for generic
locations of the $D4$-branes in $w$ they still intersect
the $D6$-branes, and hence the $4-6$ strings
giving rise to fundamental chiral multiplets
can be short. The theory also has a fully Higgsed
branch (for sufficiently large $\nf$) whose
dimension is $2\nf\nc+\nc$ and many mixed
Higgs-Coulomb branches. These branches can be
analyzed using the brane description, as in section 3.

For negative $r$ one similarly finds an
$Sp(\nc/2)$ gauge theory with $2\nf+8$
fundamental chiral multiplets and an antisymmetric
tensor. The moduli space corresponding to
giving an expectation value to the antisymmetric
tensor is $\nc/2$ dimensional. In the brane construction
it corresponds to separating pairs
of fourbranes in $w$; the fact that in this
situation pairs of $D4$-branes cannot be separated
is well known \cite{gimon}. Generically in this moduli
space $Sp(\nc/2)$ is broken to
$Sp(1)^{\nc/2}$. There is again no
Yukawa coupling between the antisymmetric
tensor and the fundamentals. The brane
and gauge theory moduli spaces can be
checked to agree as before.

\subsection{The General Case}

For generic rotation
angle $\theta$ (\ref{BST38}) the adjoint field
$\Phi$ discussed in the previous subsection is
massive \cite{bar}. Its mass $\mu(\th)$ varies smoothly
between zero at $\th=\pi/2$ and $\infty$ for
$\theta=0$. The superpotential describing this
system is
\beq
\label{eq:sup4}
W=Q\tilde{S}Q+\tilde QA\tilde{Q}+\Phi A\tilde{S}+
\mu(\theta)\Phi^2
\eeq
For non-zero $\mu$ we can integrate $\Phi$ out
and find the superpotential

\beq
\label{eq:sup5}
W=Q\tilde{S}Q+\tilde QA\tilde{Q}+\frac{1}{\mu(\theta)}
(A\tilde{S})^2
\eeq
for the remaining degrees of freedom. When $\theta\to 0$,
$\mu\to\infty$, and (\ref{eq:sup5}) approaches
(\ref{eq:sup1}). When $\theta=\frac{\pi}{2}$, the mass
$\mu$ vanishes and it is inconsistent to integrate
$\Phi$ out.

The analysis of the previous two subsections can be
repeated in this case with very few qualitative differences.
Consider \eg\  turning on a positive FI D-term $r>0$ (\ref{rd}).
Its effect is still to induce an expectation value
for the field $\tilde S$ and break $U(\nc)\to SO(\nc)$.
The only difference with the previous analysis is that
now for non zero $\langle\tilde S\rangle$
the superpotential (\ref{eq:sup5}) includes a mass
term for $A$; hence we can integrate it out and find
a quartic superpotential $W\sim (\tilde Q\tilde Q)^2$.
This is rather standard in brane theory \cite{gk} and the
analysis of the moduli space of vacua can be repeated
for this case.

As mentioned in the beginning of this section, it
is possible to construct brane configurations
in which some of the $D6$-branes are placed outside
the orientifold; they may also be rotated
with respect to it as in (\ref{BST38}).
For example, we could remove the $2\nf$ infinite
sixbranes placed on top of the orientifold before
in $\nf$ mirror pairs, leaving behind only
the eight semi-infinite sixbranes necessary for
charge conservation and anomaly cancellation.
We can then rotate each of the $\nf$ independent
$D6$-branes by an arbitrary angle $\theta_i$,
$i=1,\cdots,\nf$. Such
configurations contain $\nf$ fundamental $+$
antifundamental chiral multiplets, with a
superpotential that depends on the angles
\cite{aharony,gk}.

A particularly symmetric
configuration is obtained if we rotate the $D6$-brane
by the angle $\theta$, such that the sixbranes
near the $NS_\th$ fivebrane are parallel to it
(and their mirror images are parallel to its image).
Since this configuration has a chiral $SU(\nf)\times
SU(\nf)$ global symmetry the superpotential of
the chiral multiplets arising from these sixbranes
vanishes. Of course, the eight quarks arising
from the semi-infinite sixbranes stuck at the
orientifold are still coupled to the symmetric tensor
$\tilde S$ by the superpotential  (\ref{ntwosp}).
The analysis of this and other theories of this
sort is very similar to that described above.

It is also possible to study using branes
theories with higher polynomial superpotentials
$W=(\tilde S A)^{k+1}$ with $k>1$. Such theories
have been considered in the gauge theory context
in \cite{ils}; the basic idea for their construction
in brane theory appeared in \cite{egk,egkrs}.

\subsection{Duality}

The gauge theories constructed using
branes in the previous subsections
are known to have a dual description,
which was found in \cite{ils}.
In the gauge theory analysis it is important
to turn on the superpotential
\beq
\label{eq:sup8}
W_{\rm el}=(\tilde{S}A)^2
\eeq
to truncate the chiral ring.
The dual theory has gauge group $U(\tilde{N_c})$,
$\tilde{N_c}=6(N_f+2)-N_c$, $2N_f+8$
fundamental chiral multiplets $q$,
$2N_f$ antifundamentals, $\tilde q$, a symmetric (bar)
tensor, $\tilde{s}$, an
antisymmetric tensor, $a$, and singlet
meson fields $P$, $\tilde P$, $M_i$
($i=1,2$), with the quantum numbers
of $M_1=Q\tilde Q$, $M_2=Q\tilde S A\tilde Q$,
$P=Q\tilde S Q$, $\tilde P=\tilde Q A \tilde Q$.
Thus, $M_i$ are $2\nf\times (2\nf+8)$
matrices, $P$ is a $(2\nf+8)\times
(2\nf+8)$ matrix while $\tilde P$ is a
$2\nf\times 2\nf$ matrix.
The magnetic superpotential is \cite{ils}:
\beq
\label{eq:sup9}
W_{\rm mag}
=(a\tilde{s})^2 +M_2q\tilde{q}+M_1q a\tilde{s} \tilde{q}+ P
q\tilde{s} q+\tilde P\tilde{q} a \tilde{q}
\eeq
In brane theory $N=1$ duality arises as an
equivalence between the moduli spaces of vacua
of the electric and magnetic theories, which
can be seen by moving NS fivebranes
past each other \cite{egk,gk}. It has been
observed in many examples (although
the general statement has not been proven)
that when parallel NS fivebranes cross each
other there is a phase transition in the
low energy physics. On the other hand, when the
fivebranes are not parallel, the transition
in which they meet in space and exchange
places is smooth \cite{gk}, and gives rise to Seiberg's
duality.

In our case, $N=1$ duality is obtained by
displacing the $NS_\th$ fivebrane in $x^6$
until it meets the orientifold (with the
\nsp-brane on top of it) and its mirror image,
and then the $NS_\th$ fivebrane and its
mirror image exchange places. For $\theta=0,
\pi/2$ two or more of the branes involved
are parallel to each other and hence
there is a phase transition.
For generic $\theta$ the branes are not parallel
and one expects the transition to be smooth.
Thus the brane construction ``explains'' the
necessity to turn on the superpotential
(\ref{eq:sup8}) in gauge theory to get
a dual description.

Comparing (\ref{eq:sup5}) to (\ref{eq:sup8})
we see that the theory constructed from
branes is obtained from that of \cite{ils}
by adding to the electric superpotential
the composite operators
$Q\tilde S Q$ and $\tilde Q A\tilde Q$.
The duality of \cite{ils} predicts that
the corresponding magnetic theory has
the superpotential
\beq
\label{eq:sup10}
W_{\rm mag}=(a\tilde{s})^2+M_2q\tilde{q}+ M_1q a \tilde{s} \tilde{q}
+Pq\tilde{s} q+ \tilde P\tilde qa \tilde{q}+P+\tilde P
\eeq
The fields $P$, $\tilde P$ are massive
and can be integrated out. The F-term
vacuum equations are:
\beq
\label{eq:p}
q^{f_1}\tilde s q^{f_2}+\delta^{f_1f_2}=0
\eeq
\beq
\label{eq:q}
\tilde{q}^{g_1}a\tilde{q}^{g_2}+J^{g_1g_2}=0
\eeq
where $J$ is the invariant tensor of the
symplectic group,
$f_i=1,\cdots, 2\nf+8$ and $g_i=1\cdots, 2\nf$.
The first
equation means that the rank of the matrix
$qq$ (and, therefore, of the
matrix $q$) is $2N_f+8$,
while the second equation means that the rank
of $\tilde{q}\tilde q$ and of $\tilde{q}$ is
$2N_f$. Therefore, the magnetic group
is Higgsed, and the number of
magnetic colors decreases by $4N_f+8$, to
\beq
\label{eq:dual1}
\tilde{N_c}=2N_f+4-N_c
\eeq
The number of flavors does not decrease,
since for each quark eaten up by the Higgs
mechanism, a new
one appears from decomposing
$\tilde{s}$ and $a$ with respect
to the lower rank gauge group. The two magnetic meson
fields $M_1$, $M_2$ get a mass and decouple.

A quick way to derive the dual gauge group $U(\tilde \nc)$
(\ref{eq:dual1}) using branes is to turn on
the FI D-term $r$ and then exchange the
$NS_\th$ fivebrane and its mirror image
by moving them in $x^6$. Since this involves
manipulations in $SO(\nc)$ or $Sp(\nc/2)$
$N=1$ SQCD realized using an $O6$-plane,
we can use the results of
\cite{egkrs} for this system.
For example, for $r>0$ we have
an $SO(\nc)$ gauge theory with
$2\nf$ fundamental chiral multiplets
and exchanging the $NS_\th$ fivebrane
and its mirror image leads to a
magnetic theory with gauge group
$SO(\tilde \nc)$ given by
(\ref{eq:dual1}). A similar analysis
gives the same answer for $r<0$
using the duality for $G=Sp(\nc/2)$
with $2\nf+8$ fundamentals.
Alternatively, one can use linking
numbers \cite{hanany} in the presence
of an orientifold to determine the final
configuration.  Needless to say, this leads
to the same results.
\vskip .2in
\noindent
{\bf Acknowledgements}
\vskip .1in
\noindent
We thank J. Lykken, Y. Oz,
O. Pelc, A. Schwimmer
and S. Trivedi for discussions.
This work is supported in
part by the Israel Academy of Sciences and
Humanities -- Centers of Excellence Program.
The work of A. G. is supported in
part by BSF -- American-Israel Bi-National
Science Foundation. S. E. and A. G.
thank the Einstein Center at the Weizmann
Institute for partial support. D. K. is supported
in part by a DOE OJI grant. A. G. thanks the
EFI at the University of Chicago, where part of this
work was done, for its warm hospitality.
\vskip .2in
\noindent
{\bf Note added:} After this work was completed
we received \cite{lll} which discusses theories
related to those studied in section 4.

\end{document}